\documentclass[12pt,eqsecnum,preprint]{aastex}

\begin{document}

\title{The FRII Broad Line Seyfert 1 Galaxy: PKSJ 1037-2705}
\author{Brian Punsly\altaffilmark{1}, Tracy E. Clarke\altaffilmark{2}, Steven Tingay\altaffilmark{3}}\author{Carlos M.
Guti\'{e}rrez\altaffilmark{4}, Jesper Rasmussen\altaffilmark{5} and
Ed Colbert\altaffilmark{6}} \altaffiltext{1}{4014 Emerald Street
No.116, Torrance CA, USA 90503 and ICRANet, Piazza della Repubblica
10 Pescara 65100, Italy, brian.m.punsly@L-3com.com or
brian.punsly@gte.net}\altaffiltext{2}{Naval Research Lab, 4555
Overlook Ave SW Washington, DC 20375-5351, Interferometrics Inc.,
13454 Sunrise Valley Drive, Herndon, VA  20171, USA}
\altaffiltext{3}{Curtin University of Technology, Department of
Imaging and Applied Physics, GPO Box U1987, Perth, Western
Australia, 6102, Australia}\altaffiltext{4}{Instituto de Astrofisica
de Canarias, Via Lactea, La Laguna, E-38205 Tenerife,
Spain}\altaffiltext{5}{Observatories of the Carnegie Institution of
Washington, 813 Santa Barbara Street Pasadena, California 91101
(Chandra fellow)} \altaffiltext{6}{Physics and Astronomy Department,
Johns Hopkins University Baltimore, MD 21218}

\begin{abstract}In this article, we demonstrate that PKSJ 1037-2705 has a weak accretion flow
luminosity, well below the Seyfert1/QSO dividing line, weak broad
emission lines (BELs) and moderately powerful FRII extended radio
emission. It is one of the few documented examples of a broad-line
object in which the time averaged jet kinetic luminosity,
$\overline{Q}$, is larger than the total thermal luminosity (IR to
X-ray) of the accretion flow, $L_{bol}$. The blazar nucleus
dominates the optical and near ultraviolet emission and is a strong
source of hard X-rays. The strong blazar emission indicates that the
relativistic radio jet is presently active. The implication is that
even weakly accreting AGN can create powerful jets. Kinetically
dominated ($\overline{Q}>L_{bol}$) broad-line objects provide
important constraints on the relationship between the accretion flow
and the jet production mechanism.
\end{abstract}


\section{Introduction}It is unclear how the enormous stored energy
in the radio lobes of powerful FRII radio sources is related to the
thermal luminosity of the accreting gas that flows toward the
central black hole. On the one hand, it seems reasonable to expect
them to be unrelated, since the bulk of the lobe plasma was ejected
from the central engine $\sim 10^{7}$ years earlier than the optical
flux from the accretion flow. Therefore, one expects to find many or
most FRII sources with fossil lobes, no jets and a weak accretion
flow. Yet surprisingly, low frequency surveys such as the 3C survey
(which are biased towards sources with strong lobes, i.e. steep
spectrum radio emission) contain many sources with FR II lobes in
which high dynamic range observations reveal jets that can be traced
back to the central quasar, within the resolution of the radio
telescope ($\sim$ 1kpc for nearby sources) and many others have
strong radio cores and jets on VLBI (parsec) scales (eg, Cygnus A,
3C 175, 3C 334, 3C 215) \citep{bri94,pun01}. On the other hand,
there has been considerable research attempting to connect the
accretion state with the jet kinetic luminosity, $Q$ - with the
conflicting conclusions being the subject of much controversy
\citep{raw91,wil99,wan04,bor02}. A truly rigorous and unbiased
analysis of the connection between the accretion state and $Q$
cannot exclude all of the optically weak FRII radio sources a
priori, the narrow line radio galaxies (NLRGs). Many or all of the
NLRGs are optically obscured by dusty molecular gas, so a proper
treatment must probe inside this gas by looking at the Mid-IR
emission. Alternatively, some NLRGs might have weak central AGN and
are not significantly obscured. The Spitzer Telescope Mid-IR
analysis of a large sample of nearby 3C FRII NLRGs in \citet{ogl06}
seemed to indicate that there was a roughly equal mix of FRII NLRGs
that had no hidden quasar and those that had a hidden quasar. The
implication is that there is a weak central AGN in many of the FRII
objects, of which a large proportion have unobstructed lines of
sight to the accretion disk. Conversely, Cleary et al 2007 analyzed
an even larger sample of Spitzer observations of 3C FRII sources
(NLRGS and QSOs) and concluded that orientation effects alone
accounted for the differences in FRII NLRGs and FRII QSOs - i.e.,
for the most part NLRG had hidden QSOs of the same strength as in
FRII QSOs.
\par Another closely related debate in the literature is the
relationship of accretion luminosity, $L_{bol}$ (which is sometimes
expressed in a related form in terms of the Eddington ratio,
$R_{Edd}= L_{bol}/L_{Edd}$; where $L_{bol}$ is the bolometric
thermal luminosity of the accretion flow and $L_{Edd}$ is the
Eddington luminosity) to the jet kinetic luminosity, $Q$. For
example, \citet{bor02} concluded that a small $R_{Edd}$ is conducive
to strong jets in quasars. The study in \citet{wan04} was an attempt
to correlate various properties of the accretion flow with the jet
power. They concluded that $Q/L_{bol}$ and $L_{bol}/L_{Edd}$ were
inversely correlated in blazars. The inverse correlation claimed
between $Q/L_{bol}$ and $L_{bol}/L_{Edd}$ in \citet{wan04}, although
true, is a trivial consequence of the fact that $Q$ is very weakly
correlated with $L_{bol}$ in quasars (the subpopulation of blazars
in their sample that excluded BL-Lacs) and not unexpectedly,
$L_{bol}/L_{Edd}$ and $L_{bol}$ are strongly correlated in quasars
\citep{tin05}. Thus, it was straightforward by means of a partial
correlation analysis to demonstrate that inverse correlation between
$Q/L_{bol}$ and $L_{bol}/L_{Edd}$ is a spurious correlation
\citep{tin05}.
\par Clearly, powerful extragalactic jets occur in high accretion thermal luminosity systems, the FRII quasars. In
fact, some powerful FRII quasars have $L_{bol} > 10^{47}
\mathrm{ergs/sec}$ \citep{tin05}. The fundamental question is
whether low luminosity accretion flows can also power FRII jets. The
answer to this question will constrain the dynamics of plausible
central engines that can power FRII jets. In order to shed more
light on these matters, we have actively searched for FRII radio
sources with direct, unambiguous observational evidence of a low
luminosity accretion flow below the Seyfert 1/QSO dividing line: a
total thermal luminosity (IR to X-ray) of the accretion flow,
$L_{bol}< 2\times 10^{45}\mathrm{ergs/s}$ (which is shown in section
3.2 to be equivalent to the conventional dividing line between
Seyfert 1 and QSO broad-line objects, $M_{V}=-23$). To accomplish
this, one requires strong evidence for a direct line of sight to the
nucleus as in a nearly pole-on view, such as a blazar line of sight.
Secondly, it is preferable to have two metrics of $L_{bol}$, since
the optical/UV continuum is often highly contaminated by the high
frequency synchrotron tail of the jet emission and is at best an
upper bound to the optical/UV flux from the accretion flow. The line
strength of one or more broad UV emission lines is a more direct way
to measure the accretion flow luminosity in blazars. Previously, we
reported on the most extreme blazar in this family, 3C 216 with
$L_{bol}\gtrsim 10^{44}\mathrm{ergs/s}$ and a long term time
averaged jet kinetic luminosity, $\overline{Q}>
10^{46}\mathrm{ergs/s}$, \citep{pun07}. In fact it was shown that
$\overline{Q}> L_{Edd}$ in 3C 216. \par This article describes the
broadband properties of the core dominated radio source PKSJ
1037-2705. It was previously identified as a broadline AGN at
z=0.567 from an optical spectrum with uncalibrated flux
\citep{gut06}. The radio observations presented in section 2 of this
paper, indicate lobe emission with a 5 GHz luminosity typical of an
FRII radio source. The implication is that the time averaged jet
kinetic luminosity, $\overline{Q}\sim 5.0\times
10^{44}\mathrm{ergs/s}$ (In this paper we assume: $H_{0}$=70
km/s/Mpc, $\Omega_{\Lambda}=0.7$ and $\Omega_{m}=0.3$). In section
3, new calibrated optical observations are presented that indicate a
broad MgII emission line ($\sim 4000$ km/s), and the line strength
is consistent with a radio source that has a very low accretion flow
thermal luminosity, $L_{bol}\gtrsim 10^{44}\mathrm{ergs/s}$ (well
below the Seyfert 1/QSO dividing line). If it were not for the
powerful jet seen in a pole-on orientation, PKSJ 1037-2705 would be
a very ordinary Seyfert 1 galaxy: $L_{bol}\gtrsim 10^{44}
\mathrm{ergs/s}$.
\section{Radio Observations}The Australia Telescope Compact Array (ATCA) was used to obtain
observations of PKSJ 1037-2705 on November 28, 2002. The source was
simultaneously observed at frequencies of 4.8 and 8.64 GHz for 12
hours in the 6A array configuration. The bandwidth was 128 MHz at
both frequencies, dual polarization. The radio maps at 4.8 GHz and
8.64 GHz are shown in figure 1.
\begin{figure*}
\includegraphics[width=85 mm, angle= -0]{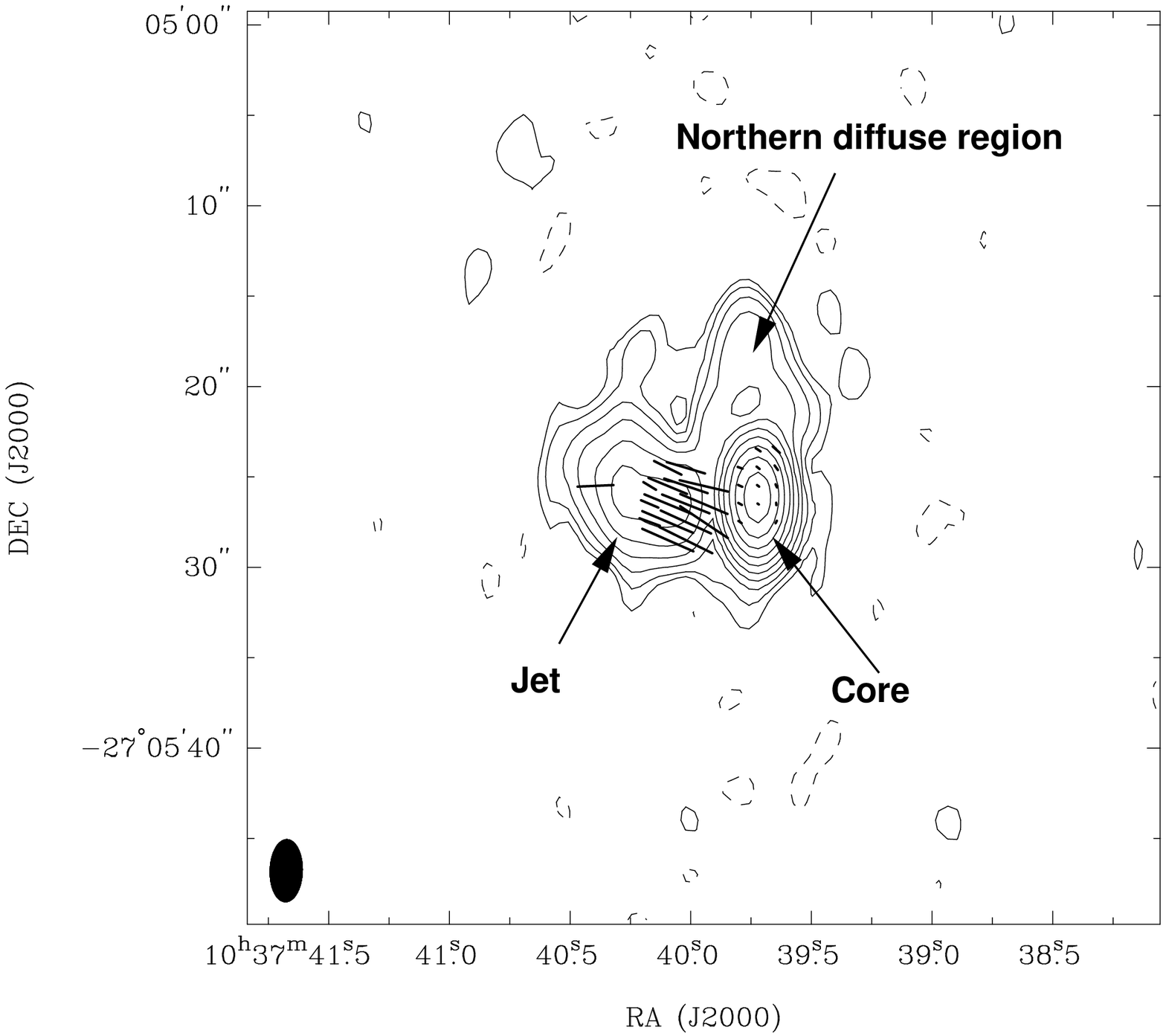}
\includegraphics[width=85 mm, angle= -0]{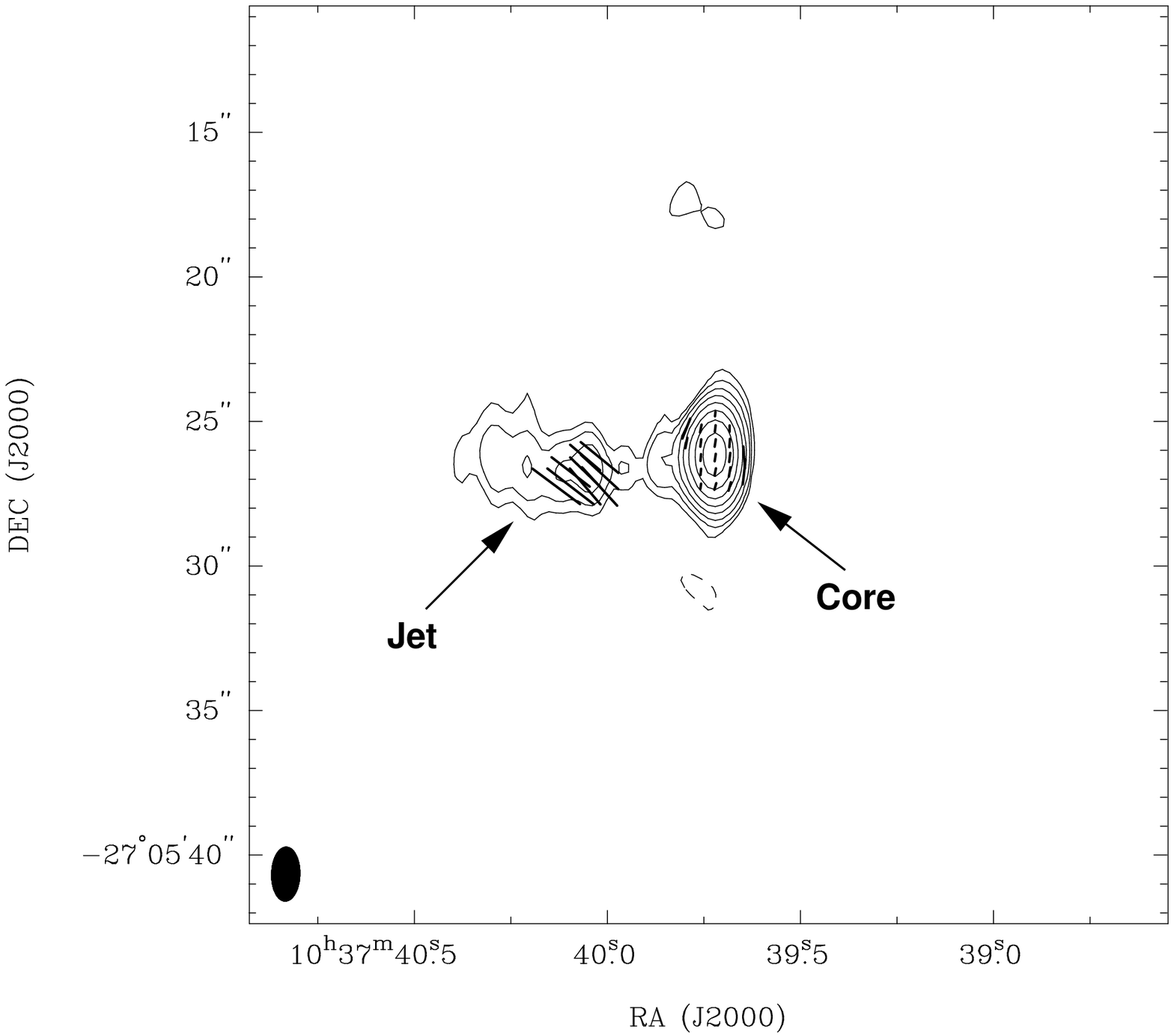}
\caption{ATCA radio maps at 4800 and 8640 MHz, from left to right.
At 4800 MHz, the beam-size is 1.75"$\times$ 3.55" at a
      position angle of $-1.5^{\circ}$.  Contour levels for the Stokes
      I emission are 1.26 mJy/beam $\times$ (-0.00125, 0.00125, 0.0025, 0.005, 0.01, 0.02,
      0.04, 0.08, 0.16, 0.32, 0.64).  The peak fractional polarization is 26.3\%. The vector lengths
      represent 5.7\% fractional polarization per arcsecond. At 8640 MHz, the beam-size is 1.03"$\times$ 1.92" at a
      position angle of $-1.5^{\circ}$.  Contour levels for the Stokes
      I emission are 1.07 mJy/beam $\times$ (-0.005, 0.005, 0.01, 0.02,
      0.04, 0.08, 0.16, 0.32, 0.64).  The peak fractional polarization is 23.9\%. The vector lengths
      represent 8.2\% fractional polarization per arcsecond.}
\end{figure*}
The components were segregated as an unresolved core, an eastern jet
and a northern diffuse region. The core flux density distribution
was modeled as a single Gaussian component. The jet and the northern
component flux densities were integrated by hand: i.e., the residual
flux density after subtracting off the Gaussian core from the image
was summed to the east (identified as the jet) and to the north. The
component fluxes in the 4800 MHz image are: core flux density =
$136.1 \pm 13.6$ mJy, jet flux density = $25.7 \pm 2.6$ mJy and the
northern diffuse flux density = $8.2\pm 0.9$ mJy. At 8640 MHz, the
component fluxes are: core flux density = $116.3 \pm 11.6$ mJy and a
jet flux density = $13.8 \pm 1.4$ mJy. The error is given by the
$10\%$ absolute flux calibration error added in quadrature to the
rms noise. Modeling the jet flux density distribution as a single
Gaussian gave a similar result. Similarly, summing the cleaned
components in the u-v plane reproduced the hand integrations in the
northern diffuse region and the jet to within a few percent. The
northern diffuse emission was not detected at 8.64 GHz. The northern
diffuse flux density in the 8.64 GHz map is $< 1.7 $ mJy (the
3$\sigma$ rms noise). Defining the radio spectral index, $\alpha$,
as $F_{\nu}\propto\nu^{-\alpha}$, we find that the core is flat
spectrum $\alpha =0.3$. The jet is extremely steep spectrum, for a
jet, $\alpha=1.1$, over 90\% of radio jets have $0.5 < \alpha < 0.9$
\citep{bri84}. Furthermore, the northern diffuse component is
unrealistically steep, $\alpha>2.5$. Clearly, significant amounts of
optically thin flux is missing due to a lack of short interferometer
spacings at 8.64 GHz (and possibly at 4.8 GHz as well). Therefore,
the jet and diffuse component spectral indices are subject to some
systematic uncertainties that we cannot easily quantify. The
complete absence of diffuse flux at 8.64 GHz to the north in a
relatively small (for a lobe) region, $\sim $ 5 arcsec across, makes
it likely that the northern component is very steep spectrum,
regardless of any systematic uncertainties. Strictly, speaking, the
eastern jet and northern diffuse component flux density should be
considered as lower limits, even at 4.8 GHz.
\subsection{A Component Model of the Radio Emission}
In order to assess the amount of diffuse flux that was missed by the
sparse array spacings, we consider the low frequency part of the
radio spectrum from archival TXS survey data.
\begin{figure*}
\includegraphics[width=105 mm, angle= -90]{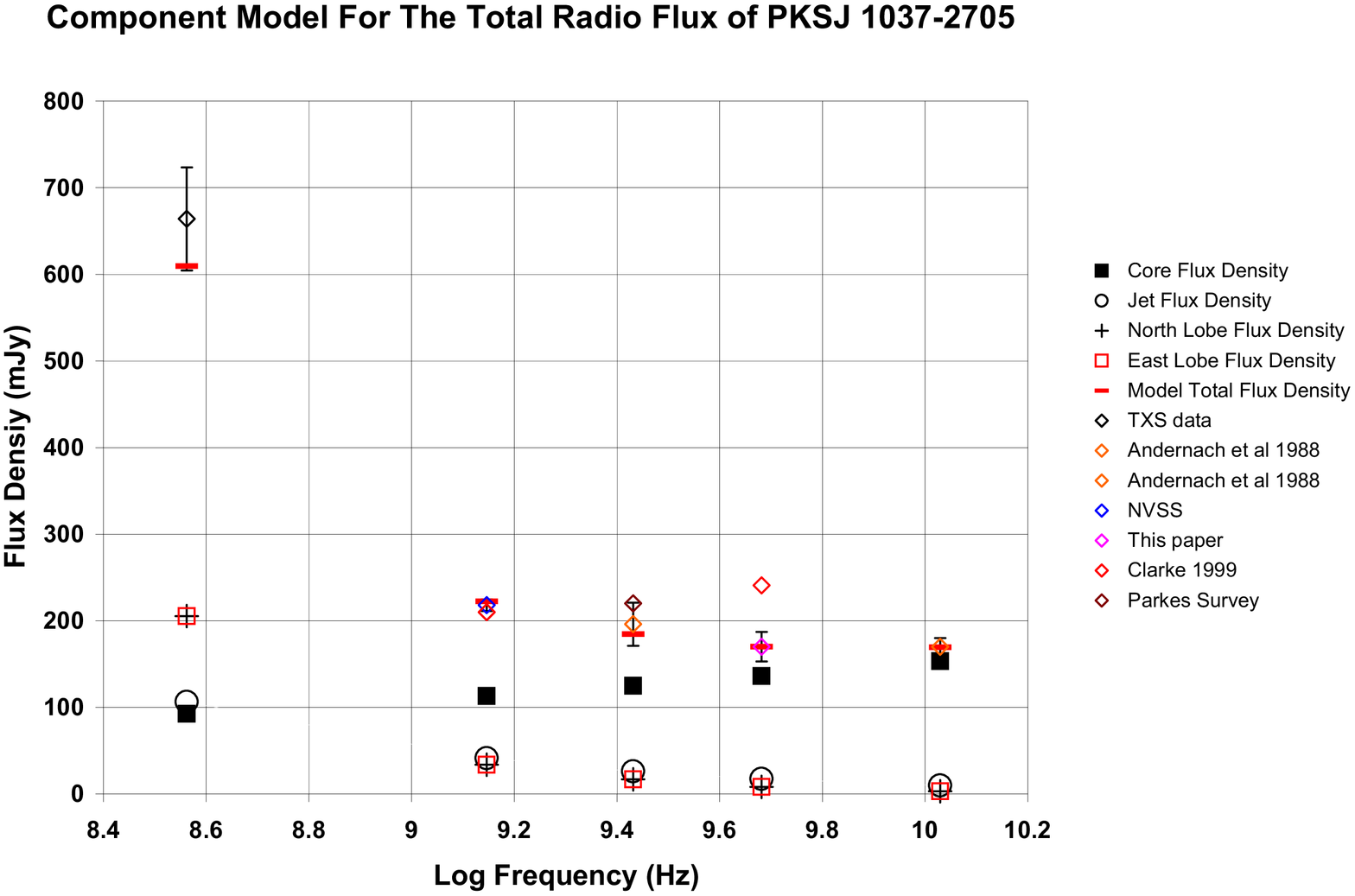}
\caption{A component model for the broadband radio spectrum of PKSJ
1037-2705. The radio data is drawn from our observations, the NASA
Extragalactic Database and archival literature. If the error bars
are smaller than the size of the marker for the observation then
they are not indicated in the figure. The model consists of a flat
spectrum radio core ($\alpha= -0.15$), a kpc scale jet ($\alpha =
0.70$) and a pair of steep spectrum lobes ($\alpha = 1.25$) with
equal flux densities (assumed bilateral symmetry, to first order).
The steep lobe spectrum seems to be required by the 1.4 GHz data
combined with the 365 MHz data}
\end{figure*}
Figure 2 is a component model that was created to explain the
broadband radio data. Before going into the details of the model,
the salient point is that there is a huge excess of low frequency
emission that cannot be explained without a significant steep
spectrum component. The figure contains all known archival data
points except the PMNJ survey point which was deleted for the sake
of visual clarity, but is discussed below. The need to produce a
detailed model of the components is rather apparent from the results
of a direct extrapolation of the 4.8 GHz data to lower frequency.
Using the two point spectral indices (from 4.8 GHz to 8.64 GHz) and
the component fluxes, one predicts a 1.4 GHz flux density for PKSJ
1037-2705 that is over 100\% larger than is observed in (see figure
2). A more sophisticated analysis is in order.
\subsubsection{The Unresolved Radio Core} First of all, the archival data clearly
indicates a variable core. We note that our previous 4.8 GHz VLA,
C-Array (relatively low resolution) data in \citet{cla99} indicated
a total flux density of 241 mJy about 40\% larger than that inferred
from our ATCA observation ($170 \pm 17$ mJy) and the archival PMNJ
data ($175 \pm 14$ mJy) in NED. Such large radio variability is
characteristic of blazars. Since the radio data is not simultaneous,
any model of the core will be inaccurate subject to the scatter
associated with the blazar variability. Clearly, the simultaneous
(to the 5 GHz) 1.4 GHz C-Array VLA data form \citet{cla99} of 210
mJy shows that the core had an inverted spectrum (contrary to our
present measurement) at that time. The relatively large flux density
at 10.7 GHz relative to 2.7 GHz flux density found by \citet{and88},
considered in conjunction with the steep spectrum lobe and jet
components, strongly suggests that the core spectrum was inverted at
that epoch as well. Thus, for the model to be an "average"
representation over time, we chose a slightly inverted spectrum,
$\alpha= -0.15$ and a flux density of 136 mJy as we measured with
ATCA.

\subsubsection{The Eastern Jet/Lobe Emission} Any reasonable choice of jet model will not affect our
conclusion that the 365 MHz data is only explained by a
predominantly steep spectrum component that is modest or small at
higher frequencies. It is curious that the jet emission is is
extremely steep spectrum. A jet spectral index above 1.0 is
extremely rare \citep{bri84}. The 4.8 GHz map in Figure 1 reveals
some excess diffuse emission to the north of the jet axis. Thus it
is likely best to think of the eastern jet as the eastern jet/lobe
emission. A significant amount of diffuse steep spectrum flux that
is co-spatially projected onto the sky plane with the jet would help
to explain the anomalously steep spectrum of the putative eastern
jet. Besides the steep spectral index, there is other evidence that
a portion of the eastern emission is from a lobe seen along the jet
axis, namely the peak of the polarization in figure 1 is not at the
end of the "linear" feature (the terminus of the linear feature is
$\approx 85-90$ kpc east of the core: there are 7.8 kpc/arsec at
this redshift for our adopted cosmology) as would be expected if the
jet terminated in a hot-spot \citep{bri94,bar90}. To the contrary,
there is virtually no detected polarized flux at the end of the
linear feature. Instead a highly polarized region is located within
$\approx 55$ kpc east of the core. The magnetic field is orthogonal
to the jet direction as is indicative of a strong knot in an FRII
jet or a terminating hot-spot coincident with the total intensity
peak, $\approx 55 - 60$ kpc to the east of the core
\citep{bri94,bar90}. Thus, the radio map is consistent with a hot
spot $\approx 55-60$ kpc from the core that terminates a pole-on
jet. The lobe might also be seen pole-on. In this interpretation,
the projection of the lobe (that is fed by the eastern jet) onto to
the sky plane in the 4.8 GHz map is a diffuse circular patch
$\approx 55$ kpc in diameter that is centered $\approx 55- 60$ kpc
slightly north of due east from of the core. The data is consistent
with this interpretation, but not definitive. In any eventuality,
the true jet flux is the total eastern flux minus the eastern lobe
flux (which we will discuss below). Approximately 40\% of kpc scale
jets have spectral indices between 0.6 and 0.7. Thus, for the
residual jet emission we picked a spectral index of 0.7 in the
composite model in figure 2. Again, any reasonable choice of jet
model will not affect our conclusion that the 365 MHz data is only
explained by a predominantly steep spectrum component that is modest
or small at higher frequencies.
\subsubsection{Lobe Emission}PKSJ 1037-2705 is likely a standard core
dominated radio source that is being viewed pole-on making the
de-projection of components complicated. The variable, flat-spectrum
core typically represents a jet beamed toward earth and the eastern
kpc-scale jet might also be beamed towards earth. The TXS 365 MHz
observation constrains the lobe flux and can be used in conjunction
with the high frequency radio maps to construct a model of the lobe
emission. The radio maps are produced from sparse array spacings so
we have likely missed extended flux at 4.8 GHz and almost certainly
at 8.64 GHz. Considering this circumstance in conjunction with our
inability to extricate lobe flux superimposed on the jet (or
co-spatial in sky plane projection with the core), the 8.2 mJy is
clearly a lower bound on the extended flux density at 4.8 GHz.
\par At a minimum, there is $8.2$ mJy of diffuse flux (to the north
of the core) in the 4.8 GHz map. The map seems to indicate a diffuse
lobe $\approx 55$ kpc in diameter. Based on the morphology of
blazars, it is not unusual for the lobe emission on the counter-jet
side to appear displaced on the sky plane roughly orthogonal to the
jet axis \citep{ant85}. We emphasize that there is likely more lobe
flux than this that is either hard to extricate from the brighter
features or was possibly missed due to the sparse array spacings. If
the source has bilateral symmetry, 16.4 mJy = $2 \times 8.2$ mJy
might be a more appropriate estimate for the lobe flux density (half
from the near lobe and half from the far lobe). The near lobe on the
jet side is not readily discernible, perhaps its projection onto the
sky plane could be confused with the powerful core or with the
bright eastern jet. A co-spatial projection onto the sky plane of an
extremely steep diffuse lobe would explain the incredibly steep jet
spectrum.
\par It is interesting to compare this conservative estimate of
16.4 mJy of 4.8 GHz extended emission with the TXS data point in
figure 2. The absence of an extended flux detection at 8.64 GHz to
the north of the radio core strongly suggests that this is very
steep spectrum lobe emission. The sources in the 3C catalog that
have steep spectrum lobes, typically have $\alpha \gtrsim 1$ between
750 MHz and 5 GHz \citep{kel69}. Using 3C radio sources as a guide,
we don't expect the spectral index from 5 GHz to 151 MHz to be any
steeper than 1.25 (see the 3C 368 data in NED). The same range
($1<\alpha<1.25$) seems to hold with the steep spectrum 7C catalog
sources \citep{wil99}. Thus, we pick $\alpha = 1.25$ for the PKSJ
1037-2705 lobe emission with 16.4 mJy at 4.8 GHz. After these
assignment of component flux densities and spectral indices, the fit
to the broad band data in figure 2 is reasonable considering the
variable core flux density and the large error in the TXS data.
There is no other way to explain the TXS data and the 1.4 GHz data
in figure 2 without a steep spectrum lobe component.
\subsection{Estimating The Jet Kinetic Luminosity}
Ostensibly, one can use the jet emission from the parsec scale radio
core to estimate, $Q$ at an epoch of emission that is within a few
years of the epoch of accretion flow emission as in \citet{cel97}.
Considering the potential variability in radio loud AGN (which could
be large if the epochs are well separated in time, especially in
blazars), simultaneous observations in all bands would be ideal for
most accurate assessment of the relation between $L_{bol}$ and $Q$.
Unfortunately, such estimates are prone to be very inaccurate. One
is observing a very small amount of dissipated energy such as X-ray
or optical emission as the powerful radio jet propagates away from
the source. One must then try to figure out the small fraction of
$Q$ that is dissipated in this region. Typically, the X-ray, radio
and optical regions are observed with different spatial resolution,
so it is unclear if one is detecting the same physical region on
parsec scales as one synthesizes the broad band data. In cases in
which there is sufficient broad band flux to make an estimate, one
is plagued with the further ambiguity of determining the Doppler
factor of the relativistic jet. This is a critical obstacle because
the luminosity from an unresolved region scales with the Doppler
factor to the fourth power \citep{lin85}. The situation is actually
worse in practice when studying blazars such as PKSJ 1037-2705 (see
the following sections for an expose of the blazar properties). The
large Doppler enhancement in blazars makes them highly variable in
virtually all bands. It is logistically very difficult to get
simultaneous broad band measurements of blazars and in fact this ia
rarely achieved. Thus, typically all the bands are sampled at
different epochs, so the error induced by the variability is imposed
on an already suspect method. More shortcomings of this method are
discussed in \citet{tin05} and a published example in which the $Q$
is apparently over-estimated, using radio core properties, by three
orders of magnitude is discussed explicitly.
\par The most accurate estimates of $Q$ should use an isotropic
estimator such as the radio lobe flux. The sophisticated calculation
of the jet kinetic luminosity in \citet{wil99} incorporates
deviations from the overly simplified minimum energy estimates into
a multiplicative factor $f$ that represents the small departures
from minimum energy, geometric effects, filling factors, protonic
contributions and low frequency cutoff. The quantity, $f$, is argued
to be constrained between 1 and 20. In \citet{blu00}, it was further
determined that $f$ is most likely in the range of 10 to 20. Thus,
choosing a value of $f=15$, \citet{pun05} converted the analysis of
\citet{wil99} to the formula in (2.1), even though it is just a time
average. This formula is an isotropic method that allows one to
convert 151 MHz flux densities, $F_{151}$ (measured in Jy), into
estimates of $\overline{Q}$ (measured in ergs/s):
\begin{eqnarray}
&& \overline{Q} \approx 1.1\times
10^{45}\left[(1+z)^{1+\alpha}Z^{2}F_{151}\right]^{\frac{6}{7}}\mathrm{ergs/s}\;,\\
&& Z \equiv 3.31-(3.65) \nonumber \\
&&\times\left(\left[(1+z)^{4}-0.203(1+z)^{3}+0.749(1+z)^{2}
+0.444(1+z)+0.205\right]^{-0.125}\right)\;,
\end{eqnarray}
where $F_{151}$ is the total optically thin flux density from the
lobes (i.e., \textbf{no contribution from Doppler boosted jet or the
radio core}). The appropriate application of this equation requires
that one must extricate the diffuse lobe emission from the Doppler
boosted core and jet. The expression, (2.1), requires 151 MHz flux
densities, so we extrapolate the 4.8 GHz data using the same lobe
flux model that worked so successfully in figure 2.  We
conservatively bound our estimate of the 151 MHz flux, by $8.2$ mJy
at 4.8 GHz with $\alpha^{4800}_{151} = 1.0$ at the low end and
$16.4$ mJy at 4.8 GHz with $\alpha^{4800}_{151} = 1.25$ at the high
end: $2.2\times 10^{44}\mathrm{ergs/sec}<\overline{Q}<9.3\times
10^{44}\mathrm{ergs/sec}$. We remind the reader that the lower bound
of 8.2 mJy at 4.8 GHz is unrealistically conservative based on
figure 2, so this is an extremely conservative lower bound.
\par Alternatively, we can use the independently derived isotropic estimator
from \citet{pun05} which is based on the assumption that the lobe
material is dominated by thermal energy and the magnetic energy
contributions are small:
\begin{eqnarray}
&&\overline{Q}=Q_{par}\approx
5.7\times10^{44}(1+z)^{1+\alpha}Z^{2}F_{151}\,\mathrm{ergs/sec}\;,\quad\alpha\approx
1\;.
\end{eqnarray}
Using the same bounds on the 151 MHz flux density as above, (2.3)
implies, $8.8\times 10^{43}\mathrm{ergs/sec}<\overline{Q}< 4.7\times
10^{44}\mathrm{ergs/sec}$. This is a conservative estimate because
it ignores any possible protonic component to the lobe energy.
Again, we note that the lower bound of 8.2 mJy at 4.8 GHz is
unrealistically conservative based on figure 2.
\par It might be a preference to use the isotropic estimators
directly without any reference to a particular model. Thus, we still
need to estimate $F_{151}$ from the 365 MHz data. Our best estimate
for the spectral index between 151 MHz and 365 MHz is the two-point
spectral index that is derived from the 365 MHz and 1.4 GHz data,
$\alpha=0.83$. This spectral index yields 1.38 Jy at 151 MHz which
translates into $\overline{Q}=8.7\times 10^{44}\mathrm{ergs/sec}$
and $\overline{Q}=4.3\times 10^{44}\mathrm{ergs/sec}$ from the
estimators in (2.1) and (2.3), respectively. The various isotropic
estimation techniques tend to indicate that $\overline{Q}\geq
10^{44}\mathrm{ergs/sec}$. Most FRII quasars in deep surveys have,
$\overline{Q} \sim 10^{44}\mathrm{ergs/sec}$, thus even the most
conservative of isotropic estimations of $\overline{Q}$ establishes
an FRII level of lobe luminosity in PKSJ 1037-2705 \citep{pun01}.
However, the only realistic estimation models of $\overline{Q}$ are
those that conform with the TXS 365 MHz data. The most reasonable
explanation of the 664 mJy data point from the TXS survey in figure
2 is that it is diffuse steep spectrum optically thin emission. One
could conjecture that it arose from a serendipitous observation of
an extremely large radio flare of the core. The required magnitude
of the relative increase in flux would rank this an an extreme flare
even by blazar standards \citep{kuh81}. If we restrict our attention
to the more reasonable scenarios in which the 365 MHz excess is not
due an extremely large flare in the core flux at low frequency then
the range of isotropic estimates in this section are $ 4.3 \times
10^{44}\mathrm{ergs/sec}<\overline{Q}< 9.3 \times
10^{44}\mathrm{ergs/sec}$.
\subsection{Summary of Radio Data Reduction} Our ATCA and the
\citet{cla99} VLA observations in conjunction with archival radio
data seems to indicate the following for PKSJ 1037-2705
\begin{itemize}
\item The radio source resides in a class of objects commonly
referred to as blazars. This deduction follows from these facts.
First, the radio map of PKSJ 1037-2705 is dominated by a powerful
unresolved radio core. Secondly, the flux density of the core shows
at least 40\% variability on the time scale of a few years. Thirdly,
the spectral index is the core is flat spectrum ($\alpha<0.5$) and
it is variable changing sign over time. All of these facts are
consistent with a relativistic jet beamed towards earth.
\item There is an excess of flux at 365 MHz that can not be
explained by the core and jet flux alone.
\item There is resolved steep spectrum lobe emission in the 4.8 GHz
map
\item The 1.4 GHz and 365 MHz data imply that the 365 MHz flux
density is comprised mainly of $>400$ mJy of very steep spectrum
emission that is most likely associated with the lobe flux.
\item The large amount of lobe flux at low frequency requires a
large amount of stored magnetic plasma energy within the lobes.
Independent isotropic estimators of this energy density indicate
that the most reasonable explanation of the lobe energy density
requires a jet with a time averaged kinetic luminosity, $ 4.3 \times
10^{44}\mathrm{ergs/sec}<\overline{Q}< 9.3 \times
10^{44}\mathrm{ergs/sec}$. - an FR II level of power.
\end{itemize}
\section{The Optical Observations}PKSJ 1037-2705 was observed with 3x1200 sec exposures using the
Telescopio Nazionale Galileo (TNG) telescope with the LR-B grism on
February 24, 2007. The spectral reduction, wavelength, flux
calibration were achieved using standard IRAF procedures (see Figure
3)\footnote{IRAF is the Image Reduction and Analysis Facility,
written and supported by the IRAF programming group at the national
Optical Astronomy Observatories (NOAO) in Tucson, Arizona.}. The
slit width was 1.5". The spectrum was smoothed with a squared box of
5 pixels. The calibration in flux was done relative to the star BD75
+325 (spectral type O5). Wavelength calibration was done with a He
lamp. The spectrum in Figure 3 is corrected for Galactic extinction
using the values given in \citet{sch98}. The most striking feature
is that the the continuum is very steep, $\alpha = 3.16$ (from 7500
\AA to 4500 \AA). The H$\beta$ emission is blocked by telluric band
absorption. The H$\gamma$ line is positioned just within the blue
side of the telluric band at 6800 \AA. However, we do clearly detect
one broad emission line (BEL), at $4399.35\AA$ that we have
identified as MgII. Our IRAF Gaussian fit yields an intrinsic MgII
line strength, $L_{MgII} \approx 1.2 \times
10^{42}\mathrm{ergs/sec}$ and the FWHM is $\approx 4000 $ km/s. The
procedure to get these values is described below.

\begin{figure}
\includegraphics[width=150 mm, angle= 0]{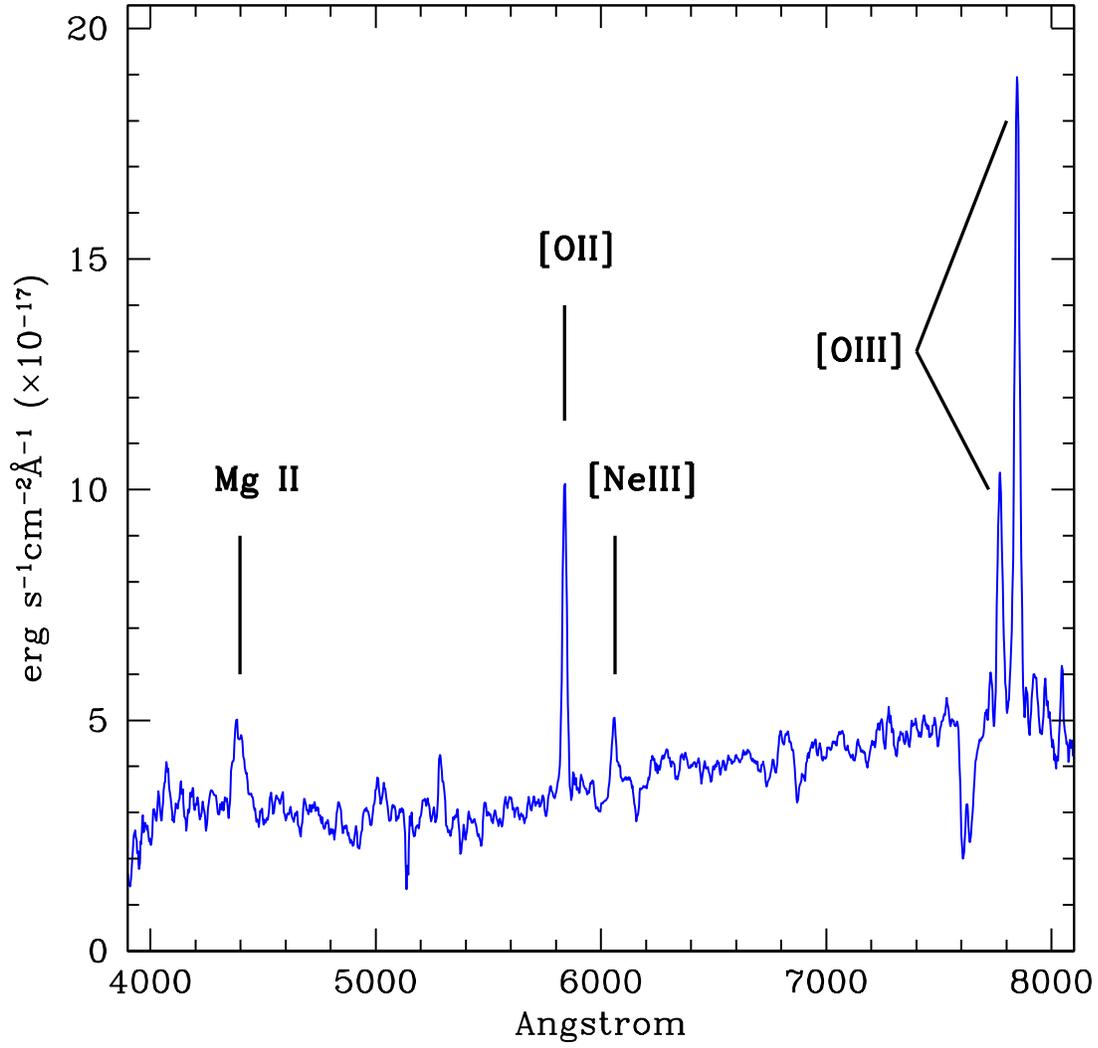}
\caption{The optical spectrum of PKSJ 1037-2705 observed with TNG
corrected for Galactic extinction. All extinction values in this
paper are from \citet{sch98}.}
\end{figure}
\begin{figure}
\includegraphics[width=150 mm, angle= 0]{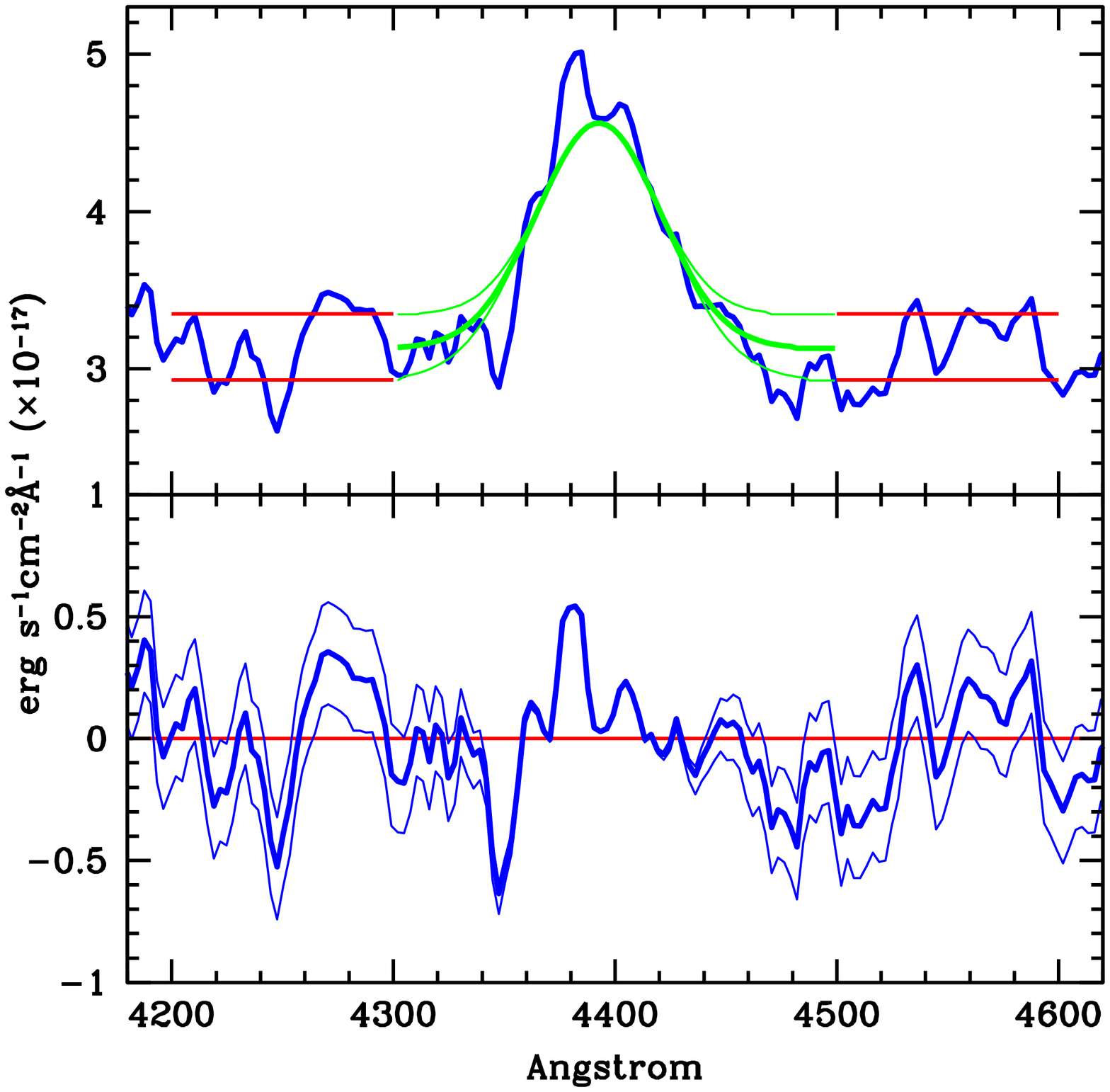}
\caption{A closeup of the TNG spectrum of PKSJ 1037-2705 in figure 3
in the vicinity of Mg II. The horizontal red lines indicate the one
sigma uncertainty in the continuum level near the Mg II BEL ($ 4200
\AA < \lambda < 4600 \AA$) caused by noise. We present a range of
Gaussian fits to the Mg II BEL as the continuum level is varied from
the average in the band to $\pm 1 \sigma$. The thick curve is the
fit to the average continuum and the thin curves are the fits to the
average continuum $\pm 1 \sigma$. The bottom frame shows the
residuals to our Gaussian fits, it looks similar to the noise in the
rest of the band.}
\end{figure}
\subsection{The MgII Broad Emission Line} The properties of the MgII
line are important to the discussion to follow. Unfortunately, the
continuum above and below the BEL is not well constrained due to
significant noise and there is apparently significant noise that
distorts the line shape near the peak. Since this measurement is
important in the following analysis, we take special care to
quantify the uncertainties in the measurement of the line strength
and FWHM.
\subsubsection{The Continuum Level} The major source of error in our estimates
arises from uncertainties in the continuum level. Since we don't
expect any other emission lines nearby, the first thing that we can
look at is the average continuum level both red-ward and blue-ward
of the BEL. The errors that are induced by the noise were obtained
by sampling the blue side of the Mg II BEL from 4200 $\AA$ to 4300
$\AA$ and on the red side from 4500 $\AA$ to 4600 $\AA$. We combined
the red and the blue bands to estimate the local continuum level as
$3.13 \pm 0.21 \times 10^{-17} \mathrm{ergs/s/cm^{2}/\AA}$. This
band of one sigma uncertainty in the local continuum level is
superimposed on the spectrum in figure 4, by two red horizontal line
segments.
\subsubsection{The Gaussian Fits} Figure 4 shows the single Gaussian fits
to the data with the continuum set at the three levels noted in the
subsection above, nominal (the average continuum level) and nominal
$\pm 1 \sigma$. A single Gaussian fit to the line with the continuum
level fixed is given by option "h" in IRAF splot. The IRAF generated
fits have residuals that look consistent with the noise level
surrounding the BEL. The FWHM values were extracted from the fitted
Gaussian models after correcting for instrumental broadening which
is about 20 $\AA$ measured from the sky and arclines. Our analysis
indicates that the three single Gaussian fits give a range of
uncertainty from the single Gaussian fit to the nominal continuum
that is expressed as $L_{MgII} = 1.23^{+.37}_{-.41} \times
10^{42}\mathrm{ergs/sec}$ and $\mathrm{FWHM}=4266^{+321}_{-570}$
km/s. If we consider a conservative 20\% uncertainty in absolute
flux calibration (this affects $L_{MgII}$, but not to the FWHM), we
compute that $L_{MgII} = 1.23^{+.69}_{-.57} \times
10^{42}\mathrm{ergs/sec}$.

\subsubsection{The Non-Gaussian Fit} A single Gaussian fit might not be
the best model of the MgII BEL in PKSJ 1037-2705. There is a
significant red-ward asymmetry and there is an irregular peak
possibly (the significant excess residual in figure 4) from a strong
narrow line component (see section 3.5 for the evidence of a strong
narrow line emissivity that is excited by the propagating jet).
Another possible reason for the distortion from a smooth profile
near the line peak is that the MgII line is actually a doublet.
Thus, we are interested in estimating $L_{MgII}$ and the FWHM
without assuming any particular parametric form (such as a single
Gaussian). The IRAF option "e" allows us to directly integrate the
flux above a certain continuum level and this seems suitable for our
purposes. Given the irregular line shape this direct approach is
preferable to the single Gaussian model for computing $L_{MgII}$.
Again, we adopt a conservative 20\% uncertainty in absolute flux
calibration and our direct integration for each of the three levels
of continua in Figure 4 yield a spread of values about the nominal
continuum estimate of $L_{MgII} = 1.24^{+.58}_{-.48} \times
10^{42}\mathrm{ergs/sec}$. Which is very close to the value and
uncertainty found from the single Gaussian fits in the previous
subsection.
\par We compute the FWHM in a similar manner with the direct
integration in IRAF option "e". We directly measured the width (in
$\AA$) of the spread in the data halfway between the peak and the
continuum. We corrected for instrumental broadening as in the
previous subsection and found a spread in the FWHM estimates
relative to the value found using the nominal continuum given by
$\mathrm{FWHM}=3504^{+313}_{-799}$ km/s. The nonparametric fit and
the Gaussian fits give values of the FWHM that are different at the
1.2 $\sigma$ level.

\subsection{Estimating $L_{bol}$ from the Continuum
Spectrum} The total bolometric luminosity of the accretion flow,
$L_{bol}$, is the thermal emission from the accretion flow,
including any radiation in broad emission lines from photo-ionized
gas, from IR to X-ray. Since we do not have complete broadband
coverage of the SED for the accretion flow, one can estimate
$L_{bol}$ by means of a composite SED. The most important samples
for the composite quasar spectra are the HST observations since
  these cover the rest frame EUV (extreme ultraviolet) region in which it was
  once believed that much of the quasar energy was hidden
  \citep{zhe97}. Figure 5 is a composite spectral energy distribution
  of a radio quiet quasar with $M_{V}=-25$ (roughly the average value
  in the \cite{zhe97} HST sample). This spectrum, in combination with
  the broad emission lines, represents the ``typical'' radiative
signature of a strong accretion flow onto a black hole in the
  absence of an FR II jet. This signature is empirical and is
  independent of all theoretical models of the accretion flow.
\begin{figure}
\plotone{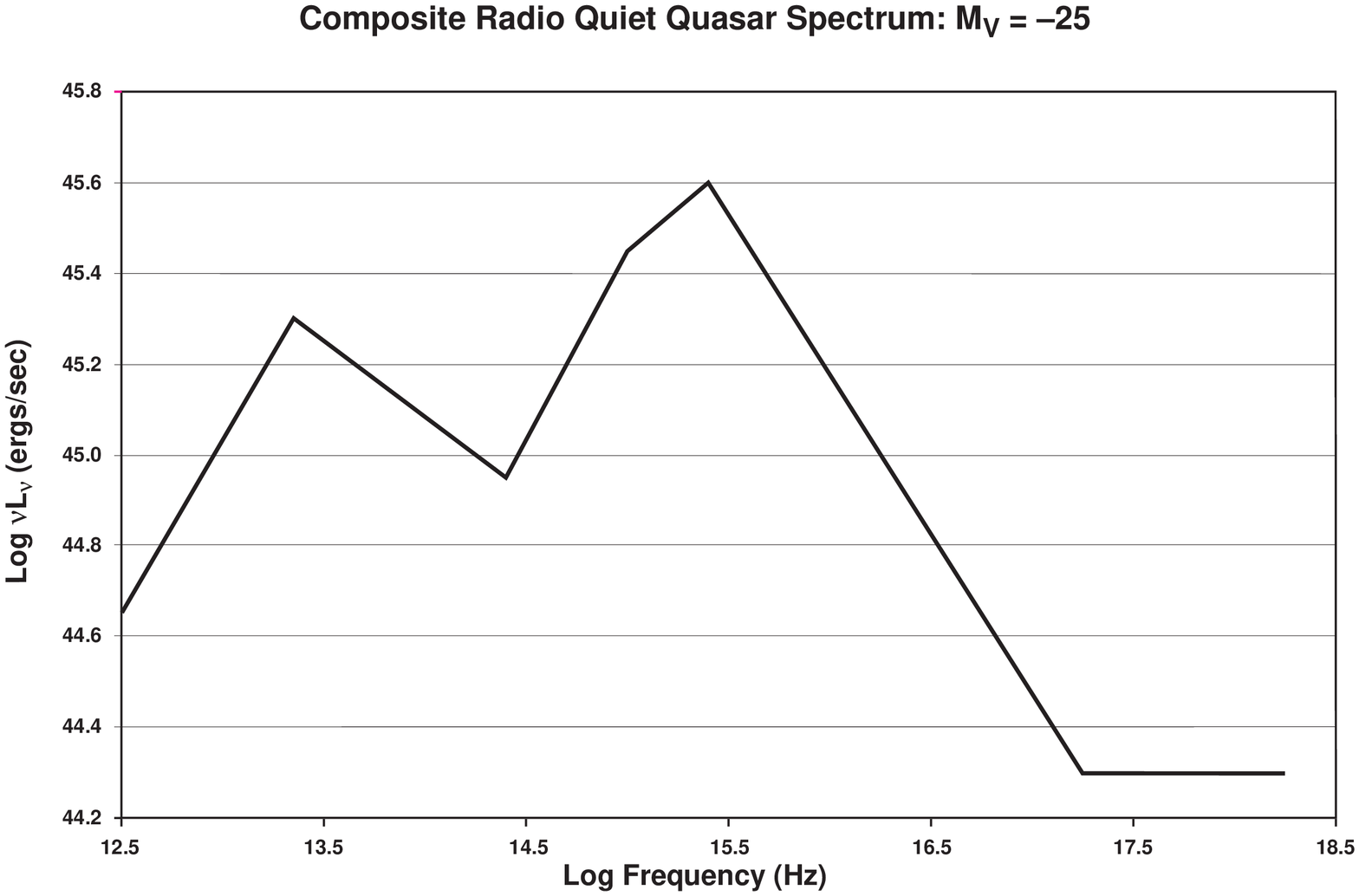}
    \caption{The composite spectral energy distribution of a quasar
      with $M_{V}=-25$ from the combined data of
      \citet{elv94,zhe97,tel02,lao97}}
\end{figure}
\par Figure 5 is a piecewise collection of power laws that approximate
      the individual bands. The IR and optical data are from the
      composite spectrum of \citet{elv94}, the NUV (near ultraviolet)
      and EUV data are from the HST composites of \citet{zhe97,tel02}
      and the X-ray portion of the composite is from \citet{lao97}. In order to compute
      the bolometric luminosity of the
      accretion flow from the composite spectral energy
      distribution, one can estimate (from the tables in \citet{zhe97}and \citet{tel02}) that 25 percent of
the total optical/UV quasar luminosity is reprocessed in the broad
line region. Thus, for this $M_{V}=-25$ radio quiet composite SED,
      $L_{bol}=1.35\times 10^{46}\mathrm{ergs/sec}$. If one also
      assumes that the shape of the composite spectrum in Figure 5 is
      independent of quasar luminosity, a simple approximate formula
      is obtained that relates the k-corrected absolute visual magnitude with bolometric
      luminosity,
\begin{eqnarray}
&& L_{bol}\approx 1.35 \times 10^{\frac{-25-M_{V}}{2.5}}\times
10^{46}\mathrm{ergs/sec}\;.
\end{eqnarray}
As a more general alternative to (3.1), if $L(\nu)_{\mathrm{obs}}$
is the observed spectral luminosity at the AGN rest frame frequency,
$\nu$, then $L_{bol}$ is estimated as
\begin{eqnarray}
&& L_{bol}=1.35\frac{\nu L(\nu)_{\mathrm{obs}}}{\nu
L(\nu)_{\mathrm{com}}}\times 10^{46}\mathrm{ergs/sec}\;,
\end{eqnarray}
where $L(\nu)_{\mathrm{com}}$ is the spectral luminosity from the
composite SED.
\par The problem with applying (3.1) and (3.2) to
PKSJ 1037-2705 is that one must subtract off the high frequency tail
of the synchrotron jet. A strong contribution from the jet that
masks the optical signature of the accretion flow seems likely given
the best fit continuum spectral index (defined in section 2 as
$F_{\nu}\propto\nu^{-\alpha}$) of $\alpha^{7500}_{4500}=3.16$ (from
7500 $\AA$ to 4500 $\AA$). In order to ascertain if the high
frequency tail of this blazar component masks the optical/near UV
flux from the accretion flow, we created a broadband SED (i.e., a
plot of $\nu L_{\nu}$ versus $\nu$) in figure 6.
\begin{figure*}
\includegraphics[width=95 mm, angle= -90]{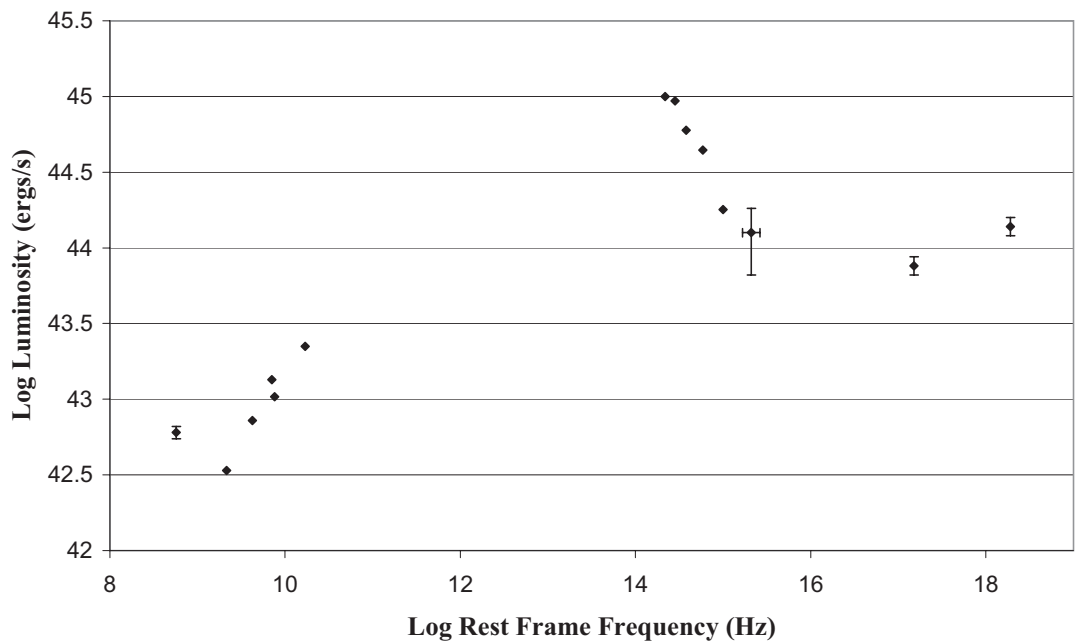}
\caption{A broadband spectral energy distribution or SED (i.e., a
plot of $\nu L_{\nu}$ versus $\nu$) for PKSJ 1037-2705. The SED is
corrected for Galactic Extinction. The radio data is drawn from
figure 2. The IR data is from the 2MASS survey and was accessed
through MAST. The optical data is from our observations. The UV data
point was from the GALEX shallow sky survey and was retrieved
through MAST. The X-ray data points come from the fit to our
Newton-XMM observation shown in Figure 8 (see section 4). If the
error bars are smaller than the size of the data point marker then
they are not shown.}
\end{figure*}
Even though the data in the SED in Figure 6 were not obtained
simultaneously (and blazars are highly variable), the basic shape is
quite revealing. First of all, notice the peak in the SED in the IR.
The peak of the SED is clearly in excess of
$10^{45}\mathrm{ergs/sec}$ in the Mid-IR. The spectrum from
$\lesssim 10^{10}$ Hz to $\gtrsim 10^{15}$ Hz is dominated by one
component with a shape typical of the synchrotron spectrum of a
blazar jet \citep{fos97}. Compare this spectrum to the signature of
the accretion disk, the SED for radio quiet quasars in Figure 5. The
thermal SED from the accretion flow is rising from $3\times 10^{14}$
Hz to $3\times 10^{15}$, the "big blue bump". Conversely, the SED
for PKSJ 1037-2705 is a steeply decreasing function in this same
frequency range, a clear distinction from a thermal origin. The rise
of the SED toward high frequency in the X-ray regime is most likely
the inverse Compton emission from this very same jet (see the next
section for details) which is another characteristics of blazar jets
\citep{fos97}. There is no evidence of a break in the synchrotron
spectral index (the optically thin portion of the synchrotron peak)
from the IR to the near UV (see section 3.3). The GALEX data has a
large error bar associated with it, so it is not too useful. All
that the GALEX data really tells us is that there is no strong
strong thermal component in the UV.
\par The SEDs in Figures 5 and 6 indicate that the
optical component from the jet is clearly much larger than that of
the accretion disk. The data (the steep power law) indicates that
down to at least $10^{15}$ Hz, the underlying continuum is still
mostly hidden by the jet flux (see also Figure 7 and the discussion
in section 3.3). Therefore, we can use Figures 5 and 6 with equation
(3.2) to create an upper bound on the accretion disk luminosity. Due
to the rapidly falling SED, the most stringent upper bound is formed
by choosing the highest frequency UV point with relatively small
errors associated with it (not the GALEX data), $\approx 10^{15}$
Hz,
\begin{eqnarray}
&& L_{bol}=1.35\frac{\nu L(\nu)_{\mathrm{thermal}}}{\nu
L(\nu)_{\mathrm{com}}}\times 10^{46}\mathrm{ergs/sec}\nonumber\\
&& \ll 1.35\frac{\nu L(\nu)_{\mathrm{synchrotron}} + \nu
L(\nu)_{\mathrm{thermal}}}{\nu L(\nu)_{\mathrm{com}}} \times
10^{46}\mathrm{ergs/sec} = 7.64 \times 10^{44} \mathrm{erg/s}\;,
\end{eqnarray}
where $\nu L(\nu)_{\mathrm{thermal}}$ is the contribution to the SED
of PKSJ 1037-2705 from the thermal accretion flow and $\nu
L(\nu)_{\mathrm{synchrotron}}$ is the synchrotron contribution to
the SED (evaluated at $\approx 10^{15}$ Hz).
\par This upper bound
is quite conservative and it is based on not finding any true signal
of the accretion flow. We want to do better than this, so we dig
deeper into the optical data set to find more evidence of the
thermal accretion flow in the next subsections.
\begin{figure}
\includegraphics[width=100 mm, angle= -90]{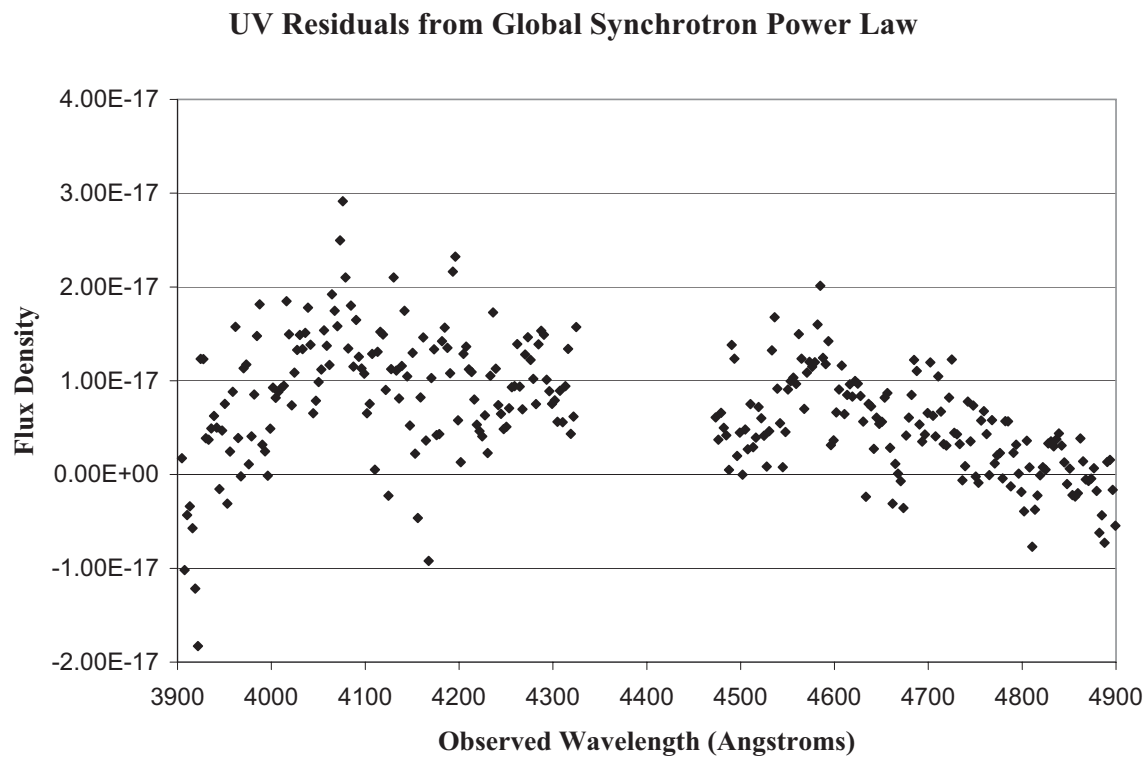}
\caption{The residual flux density of the UV (in the quasar rest
frame) continuum relative to the global synchrotron power law. A gap
is left that accommodates the Mg II BEL that significantly distorts
the continuum level. The best fit, steep power law, from 7500 $\AA$
down to 5000 $\AA$ that is given by $\alpha = 3.39$, continues to
follow the continuum very closely down to 4500 $\AA$. The figure
indicates a hint of an excess of flux below 4300 $\AA$, but the
results is not statistically significant. A slight excess would be
indicative of a weak thermal component near 4000 $\AA$. The flux
density is corrected for Galactic extinction and the units are
$\mathrm{ergs/s/cm^{2}/\AA}$}
\end{figure}
\subsection{Estimating $L_{bol}$ Based on the UV Excess from the Global Power Law}
 In order to make a better estimate than the loose upper
bound in (3.3), one can look for some weak signal that is indicative
of the thermal accretion flow. If there is no spectral ageing
effects, one expects the optically thin portion of the jet
synchrotron spectrum to be well fit by a power law. Since the
optically thin synchrotron spectrum is so steep, the best chance for
detecting evidence of a weak $L_{bol}$ is by looking for an
inflection point in the power law spectrum blue-ward of Mg II. The
farther into the UV we go, the more pronounced the thermal component
becomes, based on a typical quasar spectrum from \citet{zhe97} and
the synchrotron spectral index,
$F_{\lambda}(\mathrm{thermal})/F_{\lambda}(\mathrm{synchrotron})
\sim \lambda^{-2.5}$.
\par We note that the spectral index flattens as the end point of
the fit to the continuum on Figure 3 moves deeper into the UV. The
values of spectral index of the best fit, de-reddened spectrum for
an endpoint that goes from 5500 $\AA$ to 5000 $\AA$ to 4500 $\AA$ to
3900 $\AA$ are: $\alpha^{7500}_{5500}=3.44$,
$\alpha^{7500}_{5000}=3.39$, $\alpha^{7500}_{4500}=3.16$,
$\alpha^{7500}_{3900}= 2.92$. It appears that there is a break in
the synchrotron power law short-ward of 5000 $\AA$. This could be a
sign of a UV excess above the synchrotron power law. This is not
expected for a pure synchrotron source since spectral ageing, if
present, actually steepens the spectrum at high frequencies. A UV
excess over a pure power law fit, could be a signal of an accretion
disk starting to become noticeable in the strong glare of the
synchrotron jet. Unfortunately, the spectrum is noisy in the UV, as
noted in the last subsection, so the excess is not statistically
significant.
\par We explore this trend toward a UV excess over a power law by
first fitting the continuum spectrum from 7500 $\AA$ to 5000 $\AA$,
$\alpha^{7500}_{5000}=3.39$. We then ask the question, if this power
law were extended toward the far UV, what would the residuals of the
data look like, i.e., the excess. The results are plotted in Figure
7. There is evidence of a gradual trend for an excess extending into
the band from 4300 $\AA$ to 3900 $\AA$. We note that in the band
4900 $\AA$ to 4500 $\AA$ that the average residual flux density is
$4.5 \pm 5.1 \times 10^{-18} \mathrm{ergs/s/cm^{2}/\AA}$ and in the
shorter wavelength band, 4300 $\AA$ to 3900 $\AA$ the average excess
seems to increase to $9.1 \pm 7.0 \times 10^{-18}
\mathrm{ergs/s/cm^{2}/\AA}$. But the excess is $\sim 1 \sigma$ above
the synchrotron power law, and therefore not statistically
significant.
\par We can use this lack of statistical significance
to obtain another upper bound on $L_{bol}$. If we use the flux
excess in the band 4900 $\AA$ to 4500 $\AA$ to represent the value
at the center of the narrow band, 4700 $\AA$ (approximately 3000
$\AA$ in the quasar rest frame), then we know that the
$F_{\lambda}(\mathrm{thermal})(4700)< 1.98 \times 10^{-17}
\mathrm{ergs/s/cm^{2}/\AA}$ at the 3 $\sigma$ level. Then (3.2)
implies that $L_{bol}< 5.5 \times 10^{44}$ ergs/s at the 3 $\sigma$
level. This is a tighter upper bound than was found in (3.3).
Similarly, one can estimate the excess from the global power law at
4100 $\AA$ (2615 $\AA$ rest frame) by averaging the excess within
the band from 4300 $\AA$ to 3900 $\AA$. The flux in this band,
quoted above, inserted into equation (3.2) yields $L_{bol} =2.1 \pm
1.6 \times 10^{44}$ ergs/s, which is consistent with the upper bound
from the 4700 $\AA$ excess.
\subsection{Estimating $L_{bol}$ from
Broad Line Strengths} In order to constrain the accretion disk
thermal luminosity, a different estimator is required. The estimator
in (3.2) of bolometric luminosity of the accretion flow is a direct
measurement of an accretion flow property. Thus, it is superior to
using line emission to estimate the accretion flow power which is an
indirect estimator. The advantage of using the line luminosity is
that its strength might not be affected by the relativistic jet to
first order. However, the down side of using line luminosity is that
it is a second order indicator of accretion luminosity. One can use
either broad lines which are $\sim$ 0.01 pc from the accretion disk
or narrow lines which can occur anywhere from just outside the broad
line region out to distances of $\sim 50$ kpc \citep{bes01}. First
of all, the broad lines are closer to the photo-ionization source,
so it seems that this might be a better choice. Secondly, it has
been shown the the narrow lines can be created by jet propagation
and might not provide a reliable diagnostic of the thermal
photo-ionization source in many radio sources \citep{bes00}. This
will discussed in more detail in the next subsection. That is why
broad lines are the first choice for blazar accretion flow
estimates. The method that is most commonly used for estimating
$L_{bol}$ in blazars is to compare the line strengths to a composite
SED \citep{cel97,wan04}. Again, we note that the existence of a
radio jet tends to produce strong narrow lines in NLRGs
\citep{bes00}. Since the MgII line strength is so weak, subtraction
of the narrow line component associated with the propagating jet is
likely important to accurately estimate the broad line strength,
$L_{MgII}(BEL)$, arising from the photo-ionization of BEL clouds by
the accretion flow thermal emission. The composite NLRG ($z\sim 1$)
spectrum produced in \citet{bes01}, indicates that the MgII narrow
line strength, $L_{MgII}(NL)$ is $\approx 0.15$ of the OII line
strength. Our IRAF single Gaussian (option k) fit to the spectral
data indicate that $L_{OII}(NL)\approx 2.4\times
10^{42}\mathrm{ergs/sec}$. Thus, from the data reduction in section
3.1.3, $ 4.0\times 10^{41}\mathrm{ergs/sec}< L_{MgII}(BEL)\approx
L_{MgII} - 0.15L_{OII}(NL) < 1.5\times 10^{42}\mathrm{ergs/sec}$. We
then use the composite SED in Figure 5 combined with the HST
composite quasar line strengths of \citet{zhe97} (from which the
composite SED is derived) to compute $L_{bol} \approx 204
L_{MgII}(BEL)$, which is about 20\% more than the \citet{wan04}
formula ($L_{bol} \approx 168 L_{MgII}(BEL)$). For PKSJ 1037-2705
this relation implies $ 8.2\times 10^{43}\mathrm{ergs/sec}< L_{bol}
< 3.1\times 10^{44}\mathrm{ergs/sec}$. This is consistent with the
upper bound estimated from the optical continuum, $L_{bol}< 5.5
\times 10^{44}$ ergs/s, that was found in the previous subsection.
\subsection{Estimating $L_{bol}$ from
Narrow Line Strengths} A narrow line estimator for $L_{bol}$ could
also be tried. A seminal effort in \citet{raw91} estimated that the
total narrow line luminosity was
\begin{eqnarray}
L_{NLR} \approx 9 L_{OII3727} + 4.5 L_{OIII5007}\;.
\end{eqnarray}
For PKSJ 1037-2705 the lines strengths from the IRAF Gaussian fits
to the spectrum in figure 3 are $L_{OII3727}\approx 2.4 \times
10^{42}\mathrm{ergs/sec}$ and $L_{OIII5007}\approx 4.8\times
10^{42}\mathrm{ergs/sec}$. Thus, from equation (3.4), we estimate as
per \citet{raw91}, a narrow line luminosity of $L_{NR}\approx
4.3\times 10^{43}\mathrm{ergs/sec}$. In \citet{raw91} they picked an
arbitrary covering factor of 0.01 for the narrow line clouds (which
is likely too large by a factor of $\gtrsim 3$, see below and
\citet{wil99}). Therefore they estimated the blue bump optical/UV
luminosity, $L_{BB} = 100 L_{NLR}$. From the SED in Figure 5 and the
broad line strengths in \citet{zhe97}, $L_{bol}\approx 1.67
L_{BB}\approx 7.2\times 10^{45}\mathrm{ergs/sec}$.
\par Alternatively, \citet{wil99} claim that $L_{OIII}$ is not
reliable for the estimation of $L_{bol}$ and claim that their new
OII estimator is superior to their earlier efforts,
\begin{eqnarray}
L_{BB} \approx 5\times 10^{3} L_{OII3727}\;.
\end{eqnarray}
The \citet{wil99} estimate gives $L_{bol}\approx 1.67 L_{BB}\approx
1.9\times 10^{46}\mathrm{ergs/sec}$. The primary difference between
the \citet{wil99} and \citet{raw91} is the adhoc covering factor.
The revised \citet{wil99} value is 0.003. Note that the revised
covering factor is much more compliant with the composite quasar
spectrum of \citet{fra91} that is discussed below. With the revised
covering factor the narrow line luminosity implies that the flux
density at 3000 $\AA$ from the "big blue bump," (the ionizing
continuum) should be 30 times larger than what we observed in the
spectrum of PKS J1037-2705. This begs the question, why is the
narrow line estimator so poor for this source? Recall that
\citet{bes00} showed that $L_{NLR}$ can be from gas that was excited
by shocks that are driven by the jet in radio galaxies. By using
long slit spectroscopy, they found many regions of OII emitting gas
that are aligned with features in the jet. A study of narrow line
ratios in these aligned features indicated that the excitation state
and density was explained best by the shock models. In fact, they
showed that the smaller radio galaxies ($<120$ kpc) and those with
broad OII lines ($\sim $ 1000 km/sec) had the majority of their
narrow line gas in an ionization state that was likely excited by
jet induced shocks as opposed to photo-ionization. Thus, $L_{NLR}$
is a poor diagnostic of the thermal accretion luminosity in these
sources. PKSJ 1037-2705, is this type of source, a linear size
$<120$ kpc and a single Gaussian fit to the OII line yields a FWHM
of $\approx $ 1000 km/sec. The narrow lines are driven primarily by
the jet since the photo-ionization source associated with the
accretion disk is so weak as evidenced by the weak MgII BEL and the
low level UV continuum emission. In order to quantify how weak the
continuum ionization source is compared to the jet shock ionization
source, we note that in the composite quasar spectrum of
\citet{fra91}, (The \citet{zhe97} quasar composite does not cover
OII and OIII): $ L_{OIII}/L_{MgII} =0.1$ and $L_{OII}/L_{MgII}
=0.023$. By contrast, for PKSJ 1037-2705, we have $L_{OIII}/L_{MgII}
\approx 3.9$ and $ L_{OII}/L_{MgII} \approx 1.9$ (based on our best
fit Gaussians). It is concluded that $L_{NLR}$ is a poor predictor
of $L_{bol}$ because the ratio of the jet kinetic luminosity to the
photo-ionizing luminosity is extremely high in this quasar. In fact,
it is proposed that a potential method for selecting kinetically
dominated quasars is to find blazars (unobscured nuclei) in which $
L_{OII}/L_{MgII}> 1 $ and the MgII FWHM $>$ 2000 km/sec.

\subsection{Summary of IR/Optical Observations}In this section we
used optical observations with TNG and 2MASS archival IR data to
quantify numerous characteristics of the AGN, PKSJ 1037-2705.
\begin{itemize}
\item PKSJ 1037-2705 is blazar with a synchrotron peak luminosity
in the IR with $\nu L_{\nu}>10^{45}$ ergs/sec (see section 3.2).
\item The high frequency synchrotron tail from the blazar dominates
the optical and near UV luminosity created by the accretion flow
(see section 3.3)
\item PKSJ 1037-2705 has a Seyfert 1 thermal spectrum as evidenced
by the MgII broad emission line (see section 3.1).
\item Standard estimation techniques based on composite quasar
spectra and the broad line strength indicate that $ 8.2\times
10^{43}\mathrm{ergs/sec}< L_{bol} < 3.1\times
10^{44}\mathrm{ergs/sec}$. This is an order of magnitude below the
Seyfert 1/QSO dividing line.
\end{itemize}
\begin{figure*}
\includegraphics[width=125 mm, angle= -0]{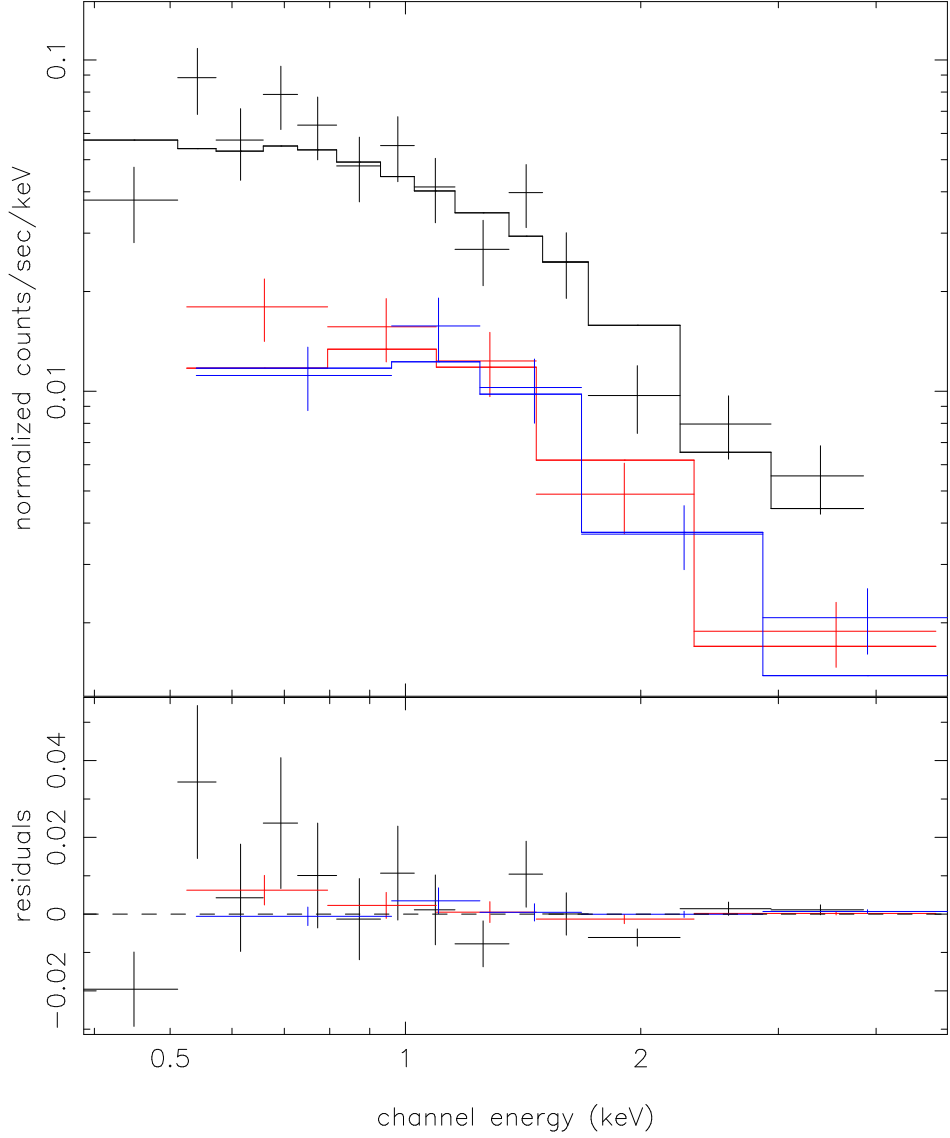}
\caption{The X-ray spectrum from XMM-Newton. The different colors
represent data for the 3 XMM-Newton cameras: black = PN, red: MOS1,
blue: MOS2}
\end{figure*}
\section{The X-ray Observations}PKSJ1037-2705 was observed on July 3, 2004 for 18 ks with
XMM-Newton. The XMM-Newton data are affected by solar flares,
yielding a cleaned exposure time of only 5 ks. Following standard
light-curve screening of the XMM-Newton data, spectra and response
products were extracted for each EPIC camera within a 30 arcsec
radius from the X-ray peak, using a surrounding 1-2.5 arcmin
point-source excised annulus for background estimates. The three
spectra were each accumulated in bins containing at least 25 net
counts and then jointly fitted in XSPEC v11.3 using standard
$\chi^2$ minimization. Various models were fitted with Galactic
absorption (Comptonized black-body, Bremsstrahlung model, accretion
disk with multiple black-body components, thermal plasma at z=0.57
and a power-law). A simple power-law absorbed by the Galaxy (ie.
with no evidence of intrinsic absorption at the source) provides the
best fit among the models that we tried. The spectrum is shown in
Figure 8. The spectral fitting was done in the 0.4-5 keV band. We
find the spectral index, $\Gamma = 1.77 \pm 0.20$ with reduced
$\chi^{2}$=1.44 for 21 degrees of freedom (the null hypothesis
probability is 0.0086). Including intrinsic absorption in the model,
the best-fit hydrogen column density of such a component is only
$\sim 20\%$ of the Galactic value and consistent with zero at 1
sigma (and with no improvement in fit quality). As a check of the
veracity of the fit, fitting was also performed using Cash
statistics on spectra binned into 5 cts/bin, but the results were
generally consistent with those of $\chi^2$ fitting reported here.
\par The absorption-corrected X-ray luminosity in the quasar rest frame (the XMM-Newton band covered in
Figure 8 transforms to 0.63-7.85 keV), is $2.49^{+0.42}_{-0.35}
\times 10^{44}$ ergs/s. This value actually is comparable to or
larger than the estimates of $L_{bol}$ that was computed from the
broadline strength, a common signature of a blazar. Archival ROSAT
data of this source (also known as IXO 37) was extrapolated with
$\Gamma = 1.7$ (similar to our value) in \citet{col02} to compute
the 2 -10 keV intrinsic luminosity. After correcting for the
incorrect redshift in \citet{col02} and extrapolating our best-fit
model to to 2 -10 keV, we find that the source was $\approx 2.5$
times more luminous at the time of the ROSAT observations. This type
of large X-ray variation (a factor of a few, in time frames of years
or shorter) is characteristic of a blazar.
\section{Discussion} In this article, we established the following
\begin{enumerate}
\item PKSJ 1037-2705 is a blazar based on radio core dominance, a
flat radio core spectrum, the large radio and X-ray variability, a
steep optical spectrum and an inordinately large X-ray luminosity
(relative to the UV and $L_{MgII}$).
\item PKSJ 1037-2705 has the extended radio flux typical of an FR II
quasar.
\item PKSJ 1037-2705 is a broadline object that is likely to have an accretion
flow luminosity an order of magnitude weaker than a quasar, i.e. it
is a Seyfert 1 galaxy.
\item PKSJ 1037-2705 is likely to be kinetically dominated based on
the spread in the estimated values of $\overline{Q}$ and $L_{bol}$
that were determined in sections 2 and 3, respectively: $4.3/3.1
=1.4 < \overline{Q}/L_{bol}< 9.3/0.82 = 11.3 $.
\end{enumerate}
\par The proof that Seyfert 1 AGN can host FRII jets is based on a few
anecdotal examples. Note that these objects are distinct from broad
line radio galaxies (BLRGs). The Mid-IR observations of
\citet{ogl06} indicate that FRII BLRGs have nuclei with quasar level
luminosity which are often obscured by dusty gas. Even though
Seyfert 1 FRII radio sources might not be exceptionally rare,
demonstrating the existence of a single object explicitly is
extremely challenging. For example, \citet{ogl06} suggest that their
weak Mid-IR subsample of NLRGs are hosted by low luminosity AGN,
i.e., Seyfert 1 level thermal luminosity. However, \citet{cle07}
would argue that the absorbing columns to these NLRGs is so dense
that even the IR can not escape and there is still a hidden quasar
buried inside.
\par The existence of Seyfert 1 nuclei in FRII sources is important
for understanding the connection between the accretion flow and the
jet power, with two possible implications. Firstly, the kpc scale
emission that was used to compute $\overline{Q}$ was ejected from
the AGN $\sim 10^{6} - 10^{7}$ years earlier than $L_{bol}$ was
created \citep{wil99,blu00}. It is not clear if there should be a
strong connection between the present value of $L_{bol}$ and the
value of $L_{bol}$ $\sim 10^{6} - 10^{7}$ years ago. On the one
hand, strong variability has been detected in the thermal emission
from Seyfert 1 nuclei with changes of a factor of 5 in luminosity in
weeks or months \citep{ant83,all85}. The observation of this type of
variability has led researchers such as \citet{all85,cut85,ant88} to
conclude that the variability time scale for $L_{bol}$ can not be
related to the thermal and viscous time scales of an accretion disk.
On the other hand, there is evidence that some high redshift quasars
have been active for $> 10^{7}$ years based on the $\mathrm{He}^{+}$
Lyman $\alpha$ "proximity effect" \citep{jak03}. The ``proximity
effect'' arises when a large bubble forms about the quasar in which
helium is highly ionized by the quasar. One can use the size of the
bubble to estimate the amount of time it would take for the
photo-ionizing source to ionize the extended region. However, one
must be cautious to extrapolate this result to much less luminous
Seyfert 1 galaxies and it does not provide evidence that the
photo-ionization source was steady within a factor of 10 during the
entire $10^{7}$ years. With all this being said, it would be a huge
stretch in reasoning to assume that in spite of the large short term
variations that have been observed in some Seyfert 1 galaxies that
$L_{bol}$ is constant to within a factor of 10 over $10^{7}$ years.
Hence, there does not seem to be any compelling observation or
underlying reason why the two quantities, $\overline{Q}$ and
$L_{bol}$ should be related, even if the accretion flow drives the
jet.
\par Alternatively, the existence of Seyfert 1 nuclei in FR II radio sources could mean that the
accretion rate is not that strongly coupled to the jet power.
Unfortunately, as a consequence of the Doppler boosting there is no
reliable contemporaneous estimate of $Q(t)$ that can be used to test
the latter hypothesis (see the discussion in section 2.2). In the
context of our previous work, we have now reported on 3 Seyfert 1
galaxies that have $\overline{Q} \geq L_{bol}$, PKSJ 1037-2705
($L_{bol}\gtrsim 10^{44}$ ergs/s), 3C 216 ($L_{bol}\gtrsim 10^{44}$
ergs/s) and PKS 1622-253 ($L_{bol}\approx 2\times 10^{45}$ ergs/s)
\citep{pun07,rod05}. All three have convincing evidence of a
contemporaneous powerful jet and a weak (Seyfert 1 level) $L_{bol}$.
The evidence for an active jet in PKSJ 1037-2705 is a powerful peak
in the synchrotron SED, $\nu L_{\nu}>10^{45}$ ergs/s, in the IR
based on archival 2MASS data. Similarly, the two epochs of X-ray
data indicates that it is plausible that the inverse Compton peak of
the SED in the X-ray band, (0.3 keV - 10 keV in the rest frame)
exceeds $10^{45}$ ergs/s during the high states (ROSAT epoch). 3C
216 has a strong peak in the synchrotron SED $>10^{46}$ ergs/s. PKS
1622-253 is one of the strongest EGRET gamma ray sources, the
gamma-ray apparent luminosity has a time average value of $\sim
10^{47}\mathrm{ergs/sec}$ and flares at $\sim
10^{48}\mathrm{ergs/sec}$ \citep{har99}. This circumstantial
evidence tends to support the interpretation that even weak (Seyfert
1 level) accretion flows onto a black hole can produce a central
engine for FRII jets. Thus, our study is supportive of the
\citet{ogl06} interpretation of the weak Mid-IR, FRII NLRG
subsample, they contain low luminosity nuclei.
\par Finally, we comment on the conjecture in \citet{bor02,mac03}
that these three Seyfert I galaxies, possessing a large jet power
(at the FRII level), are not unusual objects, based on the small
value of $L_{bol}/L_{Edd}$ in Seyfert 1 galaxies compared to QSOs
\citep{sun89}. This conclusion is a natural consequence of the claim
in \citet{bor02,mac03} that a small value of $L_{bol}/L_{Edd}$ is
conducive to jet formation. The implication is that Seyfert I
galaxies (with their relatively low values of $L_{bol}/L_{Edd}$
compared to QSOs), possessing FR II radio power, should be a common
state of AGN activity. However, this conjecture seems difficult to
reconcile with the sparsity of known FR II level extended emission
associated with broad line nuclei with Seyfert 1 level luminosity in
optically selected samples (there is not a single example in the New
General Catalog, Palomar-Green Survey or the Markarian catalog).
Such a simple proposal is also at odds with the anecdotal case of
PKS~0743$-$67 that was studied in detail in \citet{tin05} for just
this reason. PKS~0743$-$67 is an example of a quasar that has an
ultra-luminous accretion flow, $L_{bol}>2\times
10^{47}\mathrm{ergs/s}$, and has a very high Eddington rate,
$L_{bol}/L_{Edd}\approx 1$. The jet kinetic luminosity was
conservatively estimated at $\overline{Q}= 4.1 \times 10^{46}$
ergs/sec which is 2.5 times that of Cygnus A. However, this was a
conservative lower bound and the radio data also supports
$\overline{Q}\approx 10^{47}$ ergs/sec, i.e one of the most powerful
radio sources in the known Universe. Furthermore, PKS~0743$-$67 is
presently active as evidenced by the powerful ($> 1 Jy$) unresolved
VLBI radio core. A low Eddington ratio is not likely to be
determinant to FRII jet production.
\begin{acknowledgements}
The discussion of optical data is based on observations made with
the Italian Telescopio Nazionale Galileo (TNG) operated on the
island of La Palma by the Fundación Galileo Galilei of the INAF
(Istituto Nazionale di Astrofisica) at the Spanish Observatorio del
Roque de los Muchachos of the Instituto de Astrofisica de Canarias.
We are indebted to Matt Malkan and Ski Antonucci for valuable
comments that help to estimate the uncertainties in the
observational data. Tracy Clarke acknowledges that basic research in
radio astronomy at the NRL is supported by 6.1 Base funding.
\end{acknowledgements}


\begin{thebibliography}{}
\bibitem[Alloin et al (1988)]{all85} Alloin, D., Pelat, D., Phillips, M., Whittle, M, 1985,
  ApJ \textbf{288} 205
\bibitem[Andernach et al(1988)]{and88} Andernach, H. et al, 1988,
  A\& AS \textbf{73} 265
\bibitem[Antonucci (1988)]{ant88}Antonucci, R., 1988 Supermassive black holes in \emph{Proceedings of the Third George Mason Astrophysics Workshop,
Fairfax, VA, Oct. 14-16, 1986} (Cambridge University Press, New
York) p. 26-38
\bibitem[Antonucci and Cohen(1983)]{ant83} Antonucci, R. and Cohen, R., 1983,
  ApJ \textbf{271} 564
\bibitem[Antonucci and Ulvestad(1985)]{ant85} Antonucci, R. and Ulvestad, J., 1985,
  ApJ \textbf{294} 185
\bibitem[Barthel et al(1990)]{bar90} Barthel, P., Tytler, D., Thompson, B. 1990, Astron. and Astrophys. Sup. \textbf{82} 339
\bibitem[Best et al(2000a)]{bes01}Best, P., Röttgering, H., Lehnert, M. 2000 MNRAS \textbf{311} 1
\bibitem[Best et al(2000b)]{bes00}Best, P., Röttgering, H., Lehnert, M. 2000 MNRAS \textbf{311}
23
\bibitem[Blundell and Rawlings(2000)]{blu00} Blundell, K., Rawlings, S. 2000,
AJ \textbf{119} 1111
\bibitem[Boroson(2002)]{bor02}Boroson, T. 2002, ApJ
  \textbf{565} 78
\bibitem[Bridle et al(1994)]{bri94} Bridle, A. et al 1994, AJ \textbf{108} 766
\bibitem[Bridle and Perley (1984)]{bri84} Bridle, A. and Perley, R. 1984, Annu. Rev. Astron. Astrophys. \textbf{22}
319
\bibitem[Celotti et al(1997)]{cel97}Celotti, A., Padovani and Ghisellini, G. 1997, MNRAS
\textbf{286} 415
\bibitem[Clarke(1999)]{cla99}Clarke, T.E. 1999, PhD
Dissertation University of Toronto.
\bibitem[Cleary et al.(2007)]{cle07} Cleary, K.; Lawrence, C. R.; Marshall, J. A.; Hao, L.; Meier, D 2007,
ApJ \textbf{660} 117
\bibitem[Colbert and Ptak(2002)]{col02}Colbert, E., Ptak, A. 2002, ApJS
\textbf{143} 25
\bibitem[Cutri et al (1985)]{cut85} Cutri, R., Wisniewski, W.,  Rieke, G.,  Lebofsky, M.  1985,
  ApJ \textbf{296} 423
\bibitem[Elvis et al(1994)]{elv94} Elvis, M. et al 1994, ApJS
\textbf{95} 1
\bibitem[Fossati et al(1997)]{fos97}Fossati, G.; Celotti, A.;
Ghisellini, G.; Maraschi, L 1997 MNRAS \textbf{289} 136
\bibitem[Francis et al(1991)]{fra91}Francis, P. et al 1991, ApJ \textbf{373}
465
\bibitem[Gutierrez(2006)]{gut06}Gutierrez, C. 2006 ApJL, \textbf{640}, 17
\bibitem[Hartman et al (1999)]{har99}Hartman, R.. et al 1999, ApJS \textbf{123} 79
\bibitem[Jakobsen et al (2003)]{jak03}Jakobsen, P., Jansen, R., Wagner, S., Reimers, D. 2003, A \& A \textbf{397}
891
\bibitem [Kellermann et al (1969)]{kel69}Kellermann, K. I., Pauliny-Toth, I. I. K., Williams, P. J. S. 1969 ApJ \textbf{157} 1
\bibitem[Kuehr, H et al.(1981)]{kuh81} Kuhr, H., Witzel, A., Pauliny-Toth, I.I.K.,
Nauber, U. 1981, A \& AS \textbf{45}, 367
\bibitem[Laor et al(1997)]{lao97} Laor, A. et al 1997, ApJ \textbf{477} 93
\bibitem[Lind and Blandford(1985)]{lin85}Lind, K., Blandford, R.
1985, ApJ \textbf{295} 358
\bibitem[Maccarone et al(2003)]{mac03} Maccarone, T., Gallo, E., Fender, R. 2003 MNRAS \textbf{345}, L19
\bibitem[Ogle et al(2006)]{ogl06} Ogle, P., Whysong, D. and Antonucci, R.
2006 ApJ \textbf{647}, 161
\bibitem[Punsly(1995)]{pun95}Punsly, B. 1995, AJ \textbf{109} 1555
\bibitem[Punsly(2001)]{pun01}Punsly, B. 2001, \emph{Black Hole Gravitohydromagnetics} (Springer-Verlag, New York)
\bibitem[Punsly(2005)]{pun05}Punsly, B. 2005, ApJL \textbf{623} 9
\bibitem[Punsly(2006)]{pun07}Punsly, B. 2006, ApJL \textbf{651} 17
\bibitem[Punsly and Tingay(2005)]{tin05}Punsly, B., Tingay, S. 2005, ApJL \textbf{633} L89

\bibitem[Punsly et al (2005)]{rod05}Punsly, B., Rodriguez, L., Tingay, S., Cellone, S.
 2005, ApJL \textbf{633} L93
\bibitem[Rawlings and Saunders(1991)]{raw91} Rawlings, S., Saunders,
  R. 1991, Nature \textbf{349}, 138
\bibitem[Schlegel et al.(1998)]{sch98} Schlegel, D. J.,  Finkbeiner, D. P. \& Davis, M.  1998, ApJ \textbf{500}
525
\bibitem[Sun and Malkan(1989)]{sun89}Sun, W.H., Malkan, M. 1989, ApJ \textbf{346}
68
\bibitem[Telfer et al(2002)]{tel02} Telfer, R., Zheng, W., Kriss, G.,
  Davidsen, A. 2002, ApJ \textbf{565} 773
\bibitem[Vestergaard and Peterson(2006)]{ves06}Vestergaard, M. and Peterson, B. 2006, ApJ \textbf{641}
689
\bibitem[Wang et al.(2004)]{wan04}Wang, J.-M., Luo, B, Ho, L. 2004, ApJL \textbf{615} 9
\bibitem[Willott et al.(1999)]{wil99}Willott, C., Rawlings, S., Blundell, K., Lacy, M. 1999, MNRAS \textbf{309} 1017
\bibitem[Zheng et al(1997)]{zhe97} Zheng, W. et al 1997, ApJ \textbf{475} 469
\end{thebibliography}
\end{document}